\documentstyle[12pt]{article}

\newcommand{\be}{\begin{equation}}
\newcommand{\ee}{\end{equation}}
\newcommand{\ba}{\begin{eqnarray}}
\newcommand{\ea}{\end{eqnarray}}

\begin{document}
\def\pct#1{(see Fig. #1.)}

\begin{titlepage}
\hbox{\hskip 12cm ROM2F-96/35  \hfil}
\hbox{\hskip 12cm July, 1996 \hfil}
\hbox{\hskip 12cm hep-th/9607229 \hfil}
\vskip 1.4cm
\begin{center} 
{\Large  \bf  Comments \ on \ Gepner \ Models}
\vskip 24pt
 {\Large  \bf  and \ Type I \ Vacua \ in \ String \ Theory\footnote{Work
supported in part by E.E.C. Grant CHRX-CT93-0340.}}
 
\vspace{1.8cm}
 
{\large \large Carlo Angelantonj$^{\dagger}$, Massimo Bianchi$^{*}$, 
Gianfranco  Pradisi$^{*}$}
\vskip 12pt
{\large \large Augusto Sagnotti$^{*}$ and Yassen S. Stanev$^{*}$\footnote{I.N.F.N.  Fellow,
on Leave from Institute for Nuclear Research and Nuclear Energy, Bulgarian
Academy of Sciences, BG-1784 Sofia, BULGARIA.}}
\vspace{0.5cm}

{$^{\dagger}$ \sl
Dipartimento di Fisica,  Universit{\`a} dell' Aquila \\
I.N.F.N. - L.N.G.S.\\
Via Vetoio \ \ 67010 Coppito (AQ), \ \ ITALY}
\vspace{0.3cm}

{$^{*}$ \sl Dipartimento di Fisica, \ \
Universit{\`a} di Roma \ ``Tor Vergata'' \\
I.N.F.N.\ - \ Sezione di Roma \ ``Tor Vergata'', \\
Via della Ricerca Scientifica , 1 \ \
00133 \ Roma, \ \ ITALY}
\end{center}
\vskip 1.2cm

\abstract{We construct open descendants of Gepner models,  
concentrating mainly on the six-dimensional case, where they give
type I vacua with rich patterns of Chan-Paton symmetry breaking and
various numbers of tensor multiplets, including zero.
We also relate the models in $D < 10$ without
open sectors, recently found by other authors, to the generalized
Klein-bottle projections allowed by the crosscap constraint.}
\vfill
\end{titlepage}
\makeatletter
\@addtoreset{equation}{section}
\makeatother
\renewcommand{\theequation}{\thesection.\arabic{equation}}
\addtolength{\baselineskip}{0.3\baselineskip} 

\section{Introduction}

Superstring propagation on manifolds of $SU(n)$ holonomy requires extended
world-sheet superconformal symmetry \cite{wsscs}. Gepner has shown how to
describe such vacua using
tensor products of $N=2$ superconformal minimal models \cite{gep}. The
underlying Calabi-Yau (CY) manifolds are complete intersections in
weighted projective space \cite{wps}.
In principle, mirror symmetry allows the computation of world-sheet
instanton corrections \cite{wsic} to the low energy effective lagrangian. 
More recently, second-quantized mirror 
symmetry \cite{fhsv} has opened the way to computations of
non-perturbative string corrections for $N=2$ heterotic - type II dual 
pairs \cite{kv}.
The extension of these results to $N=1$ models calls for a better understanding
of the relation between heterotic and type I vacua. The conjectured 10D
duality between the $SO(32)$ heterotic string and the type I superstring 
\cite{dyn} has passed several tests \cite{test}, and
is expected to persist, in some non-naive form, after
compactification \cite{vwdual}. In this respect it is worth noticing that
the relation between type I and heterotic dilatons depends on the
space-time dimension $D$ according to \cite{wroma2}
\be
\phi_{H} = {6-D \over 4} \ \phi_{I} \ - \ {D-2 \over 16} \ \log \det G_I \quad ,
\ee
where $G_I$ is the internal metric in the type I string-frame. For instance, 
in six dimensions the heterotic dilaton is related to
internal gravitational modes of the type I string, rather than
to the type I dilaton \cite{blpssw}. 

In the past, type I models have been studied to a lesser extent than
models of oriented closed strings.
The initial proposal \cite{cargese} of identifying open-string
theories as {\it parameter-space orbifolds} of left-right symmetric theories 
of oriented closed strings has been 
brought to a consistent systematization \cite{bs,bpstor}, realizing 
Chan-Paton (CP) symmetry breaking at the microscopic level. 
The first consistent 6D
$N=1$ chiral open-string models \cite{bs} differ markedly from
perturbative heterotic K3 compactifications \cite{orbifold,gep,eoty}, 
since they include different numbers of tensor multiplets
that take part in a generalized Green-Schwarz (GS) mechanism \cite{tensor}. 
Recently, additional instances of $N=1$ type I 6D models 
have been constructed as toroidal
orbifolds \cite{gpdp,gjdp}, along the lines of \cite{ps}. A nice geometrical
setting for all these 6D models is provided by the F-theory proposal 
of \cite{vafaf},
where non-trivial scalar backgrounds allow for an effective 12D dynamics.
The variety of 6D models with different numbers of tensor multiplets may 
then be related to a corresponding variety of compactifications on 
elliptically-fibered CY threefolds \cite{morvaf}.

In this letter we discuss descendants of
type IIB models based on rational superconformal field theories. 
We mainly concentrate on 6D models, where a precise 
comparison with orbifold compactifications can be drawn. This
allows a complementary view of several known results, while introducing
a novelty, models with {\it no} tensor multiplets at all.
Here we discuss only the simplest Gepner models, related to free conformal 
field theories. We also relate the models without an open sector 
of \cite{gjdp} to the ``crosscap constraint'' of \cite{fps,pss}.
This consistency condition
for conformal field theories on open and unoriented surfaces allows, 
in general,
multiple choices for the Klein-bottle projection of the closed spectrum. 
For minimal and $SU(2)$ WZW models, this was
discussed in \cite{pss}. For six-dimensional models, it retrieves the
unique anomaly-free $N=1$ spectrum with 9 tensor multiplets and 12 
hypermultiplets found in \cite{gjdp}.
More complicated models, including $N=1$ type I models in four dimensions, 
can be analyzed in a similar fashion and are full of interesting surprises. 
For instance, 
starting from the $Z$ orbifold, one can build chiral $N=1$ type I vacua in D=4
\cite{wroma2}, where the moduli space of
untwisted closed-string scalars, $Sp(8)/SU(4)\times U(1) $, is strongly
suggestive of a twelve-dimensional interpretation \cite{bfpss}.
A more complete analysis of rational superconformal models and of some
of their deformations will be presented elsewhere \cite{abpss}.

\section{Open Descendants of Gepner Models}

In addition to the Virasoro generators $L_n$, the $N=2$ superconformal algebra 
includes two supercurrents $G^{\pm}_r$ and a $U(1)$ current $J_n$. It
admits an automorphism, known as spectral flow, under which  
\be  
L_n\rightarrow L_n +\eta J_n + {\eta^2 c\over 6} \delta_{n,0} \ ,\qquad
J_n\rightarrow J_n +{\eta c\over 3}\delta_{n,0} \ ,\qquad 
G^{\pm}_r \rightarrow G^{\pm}_{r\pm\eta} \quad ,
\ee
where $0\le \eta <1$ and $c$ is the central charge, that connects different
sectors of the spectrum.
In models describing superstring propagation, the coupling to the $N=1$
world-sheet gravitino induces a restriction to the Ramond (R)
($G$-periodic) and Neveu-Schwarz (NS) ($G$-anti-periodic) sectors.
The primary fields of the $N=2$ superconformal algebra
are labeled by the conformal weight $h$ and the $U(1)$ charge $q$. 
In the NS sector, the massless vertex operators involve
chiral primary fields with  $h=q/2$. Their supersymmetric partners in
the R sector, with $h=c/24$, arise from the spectral flow with
$\eta =1/2$.  The unique R descendant of the identity ($h=0$) is
identified with the target-space supercharge.

Gepner has shown how to construct string vacua with spacetime supersymmetry
tensoring $N=2$ superconformal minimal models \cite{gep}, that
form a discrete series with central charge $c_k = 3k/(k+2)$. For $d=(D-2)$ 
non compact (transverse) dimensions, the
total internal central charge is
$c_{I} = \sum_i c_{k_i} = 12-3d/2$, where $c_{k_i}$ are the central charges
of the various factors. The correct periodicity of the 
world-sheet supercurrent is ensured if characters from the NS 
and R sectors are combined separately. Target-space supersymmetry
follows if the supercharge has a local operator product expansion with all
vertex operators. The required truncation ($G$-projection), induced by
the restriction to integral $U(1)$ charges, is consistent with 
modular invariance if $c_{I} = 3n$, and
the ``orbifoldized'' tensor product of $N=2$ minimal models
then describes superstring propagation on a manifold of $SU(n)$ holonomy.

The general procedure for constructing type IIB vacua, described in \cite{eoty}, starts with the identification 
of the graviton orbit in the NS sector, $N^{+}_{0}$.
The other massless ($m$) and massive ($M$) NS orbits, $N^{+}_{\alpha}$,
$\alpha=1,\ldots,n_m+n_M$, are exposed by
a modular $S$ transformation on $N^{+}_{0}$. Their projections 
$N^{-}_{\alpha}$
are obtained by a $T$ transformation, while the R orbits $R^{+}_{\alpha}$ are 
obtained by an $S$ transformation from $N_{\alpha}^{-}$. Finally, a spectral
flow with $\eta =1$ determines the projected R orbits, $R^{-}_{\alpha}$.
Introducing the supersymmetric characters
\be
X_\alpha =
{1\over 2} \biggl( \
V_d (N^{+}_{\alpha} + N^{-}_{\alpha}) + 
O_d (N^{+}_{\alpha} - N^{-}_{\alpha}) -
C_d (R^{+}_{\alpha} - R^{-}_{\alpha}) -
S_d (R^{+}_{\alpha} + R^{-}_{\alpha}) \
\biggr) \quad ,
\ee 
where $\{ V_d , O_d , C_d , S_d\}$ are level-one $SO(d)$ characters, a
modular invariant partition function with space-time supersymmetry 
can be written as
\be
{\cal T}_{susy} =  \sum_{\alpha=0}^{n_m+n_M} 
\ell_\alpha |X_\alpha |^2 \quad ,
\label{tsusy}
\ee
where $\ell_\alpha = S_{0\alpha} / S_{\alpha 0}$ are the multiplicities
of each orbit. 
In order to construct open descendants of L-R symmetric closed models
as in \cite{bs}, 
one has to resolve the fixed-point ambiguity in the definition of the
characters.
The different modular invariant torus amplitudes may 
be related to the propagation on different complex manifolds, consistently with
the ADE classification of simple singularities \cite{wps}.

\section{D=8 and Other Toroidal Compactifications}

Gepner models with $c=3$ describe
compactifications on rational tori. For instance, the orbifoldized
tensor products $(k=1)^3$ and $(k=1)\times (k=4)$ correspond to the $SU(3)$
torus, while $(k=2)^2$ corresponds to the 
$SU(2)\otimes SU(2)$ torus \cite{gep}. 
The resulting open descendants
are among the rational toroidal compactifications discussed in \cite{bpstor}.
There it was shown how 
a quantized background for the NS-NS antisymmetric tensor
reduces the size of the CP group and how CP
symmetry breaking may proceed
via Wilson lines. This setting plays a crucial role in establishing 
a correspondence between F-theory on K3 and heterotic string on $T_2$
\cite{sen}. 

In general, the Klein bottle amplitude allows
for the introduction of signs $\epsilon_i = \pm 1$ in the projection of the closed-string
spectrum. This was discussed in detail for $SU(2)$ WZW models in \cite{pss}.
In a generic $r$-dimensional toroidal compactification at radii $R_i$, with
$i=1,\ldots,r$, it is possible to choose a Klein bottle amplitude 
(neglecting irrelevant factors)
\be
{\cal K} =\prod_{i=1}^{r}
 \left(\sum_{m_i\ {\rm even}} q^{
({m_i\over R_i})^2}
+ \epsilon_i \sum_{m_i\ {\rm odd}} 
q^{({m_i\over R_i})^2} \right) \quad , 
\label{exotic}
\ee
where the conventional choice is $\epsilon_i = 1$.
When at least one of the $\epsilon_i$ equals $-1$, 
there are no massless tadpoles, and thus one can not introduce 
boundary states and open strings. From a microscopic viewpoint, the
consistency of this choice may be justified from the crosscap constraint
of \cite{fps,pss}, that indeed, for a one-dimensional torus, leaves
only one relative sign between even and odd momentum sums.  
Notice that all choices of signs in (\ref{exotic}), that determine the
(anti)symmetrization of the states, respect the fusion rules.

This procedure may be generalized to
any rational model with $Z_2$ automorphisms, where the
Klein bottle projections that forbid the introduction of
the open sector do not involve massless characters
in the transverse channel. In the next section, we shall exploit this
possibility to recover the results of \cite{gjdp}, but we should mention
the simplest instance, toroidal compactification to four dimensions, 
that gives a string setting to $N=4$ supergravity coupled to
six vector multiplets.

\section{Six-Dimensional Models}

In six dimensions there are several possible types of Gepner models \cite{gep}. 
They correspond to toroidal
compactifications, to orbifolds of rational tori and
to some interacting rational $N=2$ superconformal field 
theories\footnote{More precisely, in this case the $N=2$ superalgebra extends to
an $N=4$ superalgebra, that includes an $SU(2)$ subalgebra.}.
The starting point is a ``parent'' type IIB theory, whose chiral spectrum is
uniquely fixed by target-space $N=(2,0)$ supersymmetry. 
Indeed, aside from (non-chiral) models with extended supersymmetry, that
correspond to rational points of toroidal compactifications,
any modular invariant torus amplitude
results in a massless spectrum including the $N=(2,0)$ supergravity
multiplet (the graviton, 5 self-dual tensors 
and  2 right-handed ($R$) gravitini) and 21 tensor multiplets 
(one anti-self-dual tensor, 5 scalars and 2 left-handed
($L$) tensorini). The scalar fields of the resulting low-energy
supergravity parametrize the coset $SO(5,21)/SO(5) \times
SO(21)$.

In the open descendants supersymmetry is reduced to $N=(1,0)$, and
the unoriented closed spectrum consists of the $N=(1,0)$
supergravity multiplet (the graviton, a self-dual tensor and a
$R$ gravitino) coupled to
$n^{c}_{T}$ tensor multiplets (an anti-self-dual tensor, a $L$ tensorino and
a scalar) and 
$n^{c}_{H}$ hypermultiplets (four scalars and a $L$ hyperino). The open
unoriented sector contains 
$n^{o}_{V}$ vector multiplets (a vector and a $R$ gaugino) and $n^{o}_{H}$
charged  hypermultiplets.
It should be appreciated that $n_{T}^{c} + n^{c}_{H}$
is fixed to be 21 since the Klein-bottle projection simply halves the fermi
degrees of freedom. 
The different models that we describe give rise to different values of  both
$n^{c}_{T}$ and $n^{c}_{H}$, while the presence of a self-dual
tensor in the $N=(1,0)$ supergravity multiplet leaves a net number of
$n^{c}_{T} -1$
anti-self-dual tensors. The tensor fields that flow in the
transverse channel and correspond to unphysical R-R scalars take part in a
generalized GS mechanism \cite{tensor}.

\subsection{The $(k=2)^4$ Models with $Z_2$ Symmetry}
 
The first class of models that we discuss is 
obtained tensoring four copies of the $k=2$ superconformal model.
It corresponds to the $Z_2$ orbifold of the maximal torus of
$SO(8)$ \cite{eoty}. Each $k=2$ minimal model with $c=3/2$ is
equivalent to the direct product of the Ising model with $c=1/2$ and a free
boson ($c=1$) theory at $R=\sqrt{8}$. Denoting by $\{ o,\psi ,\sigma\}$ the
three  characters of the Ising model and by $\rho_m$, $m=-3,\ldots ,4$, the
eight characters of the $c=1$ theory one finds that 
the $N=2$ characters in the
NS sector are given by $(o,\psi ) \times (\rho_{2p})$
and $\sigma \times (\rho_{2p +1})$ while those in the
R sector are given by $(o,\psi ) \times 
(\rho_{2p+1})$
and $\sigma \times (\rho_{2p})$.
The graviton orbit, that gives rise to the identity character, reads
\be
N^{+}_{0} = (o \rho_0 + \psi \rho_{4} )^4 + (o \rho_2 + \psi \rho_{-2} )^4
+ (o \rho_{-2} + \psi \rho_{2} )^4 + (o \rho_4 + \psi \rho_{0} )^4 \quad .
\ee
A proper definition of the characters requires the resolution of a fixed-point
ambiguity. There are 5 more massless ($h=1/2$) and 10 massive ($h=1$) characters
in one-to-one correspondence with those of the first class of models in
\cite{bs}. In this case, the $P= T^{1/2} S T^2 S T^{1/2}$ modular transformation 
that relates direct and transverse  M\"obius channels \cite{bs}
is diagonal. In particular, $P_{11}=-1$, and thus only 
symplectic CP groups are present in the open-string sector. This is in marked
contrast with the results of \cite{gpdp,gjdp}. 
In order to switch to unitary or orthogonal CP groups, one has to introduce
discrete Wilson lines \cite{bs}, relative phases between crosscap and boundary
operators, that break in part the internal symmetry \cite{abpss}. 
The CP symmetry is reduced, since a non-vanishing NS-NS 
antisymmetric tensor is present in the $SO(8)$ torus \cite{bpstor}.

An exhaustive analysis of all modular invariant torus amplitudes gives
three type IIB parent theories. The 
spectra of the corresponding open descendants are listed in table 1 for  
simple choices of CP multiplicities consistent with tadpole cancellation.
Non standard Klein projections,
with only massive characters in the transverse channel, are allowed in all
these cases, and give rise to the anomaly free spectrum with
$n^{c}_{T}=9$ and $n^{c}_{H}=12$ with no open-string states.  This should
be contrasted with conventional $B_{16}$ models, where massless tadpoles
require that the same closed spectrum be accompanied by anomaly-free open
sectors.


\begin{table}
\begin{center}
\begin{tabular}{|c|c|c|c|c|c|c|} 
\hline
Mod. & $n^{c}_{T}$ & $n^{c}_{H}$ & CP Gauge Group & 
$n^{o}_{V}$ & Charged Hypermultiplets & $n^{o}_{H}$ 
\\
\hline
\hline
${\bf A}_{16}$ & 7 & 14 & $Sp(4)^{\otimes 4}$ 
& 40 & $({\bf 4},{\bf 1},{\bf 4},{\bf 1})
\oplus ({\bf 1},{\bf 4},{\bf 1},{\bf 4})$ & 96
\\
 & & & & & $\oplus 2 ({\bf 4},{\bf 1},{\bf 1},{\bf 4}) \oplus
2 ({\bf 1},{\bf 4},{\bf 4},{\bf 1})$ & 
\\
\hline
${\bf B}_{16}$ & 9 & 12 & $Sp(4) \otimes Sp(4)$ 
& 20 & $({\bf 10},{\bf 1})\oplus ({\bf 1},{\bf 10})$ & 20
\\
\hline
${\bf D}_{16}$ & 5 & 16 & $Sp(8)^{\otimes 4}$ 
& 144 & $({\bf 8},{\bf 8},{\bf 1},{\bf 1})
\oplus ({\bf 1},{\bf 1},{\bf 8},{\bf 8})$  & 256
\\
 & & & & & $\oplus ({\bf 8},{\bf 1},{\bf 8},{\bf 1})\oplus
({\bf 1},{\bf 8},{\bf 1},{\bf 8})$ &  
\\
\hline
\end{tabular}
\end{center}
\caption{Open descendants of $(k=2)^4$ $Z_2$ Gepner model.}
\end{table}

\subsection{The $(k=2)^4$ Models with $Z_4$ Symmetry}

Another class of rational models arises from a $Z_4$ orbifold of the
$SU(2)^{\otimes 4}$ torus\footnote{At the closed-string level, this
rational model is equivalent to the 
$Z_2$ orbifold of the $SO(8)$ torus
\cite{eoty}, but for the absence of a NS-NS antisymmetric tensor 
background.}.   
The type IIB partition function may be written in terms of 64
characters, and corresponds to the second class of models in
\cite{bs}. In addition to the identity character, there is another massless
self-conjugate character,
16 complex massless characters, 2 massive self-conjugate characters with
$h=3/2$, two sets of 8 massive characters each with
$h=3/4$ and $h=5/4$, respectively, and, finally, 12 self-conjugate and 16
complex massive characters with $h=1$.

An exhaustive analysis of all modular invariant torus amplitudes gives
rise to seven $N=(2,0)$ type IIB parent theories. The 
spectra of the corresponding open descendants are listed in table 2 for 
simple choices of CP multiplicities consistent with tadpole cancellation.
The diagonal (${\bf D}_{64}$) and charge-conjugation
(${\bf C}_{64}$) modular invariants give rise to quite  different descendants.
The ${\bf D}_{64}$ model with standard Klein bottle projection 
yields a closed-string spectrum with one tensor multiplet
and 20 hypermultiplets, while the ${\bf C}_{64}$ model leads to
9 tensor multiplets and 12 hypermultiplets. Similar results have been
found in 
\cite{gpdp,gjdp}. In the ${\bf C}_{64}$ case, although the 
closed spectrum
is anomaly free, one has to introduce open strings to cancel unphysical
tadpoles in the transverse channel. The resulting CP group is at most
$Sp(16)\otimes Sp(16)$ for the ${\bf D}_{64}$ 
case and $U(8)\otimes U(8)$ for the
${\bf C}_{64}$ case. The introduction of discrete Wilson
lines \cite{bs} leads to CP group enhancement
to $U(16)\otimes U(16)$ for the ${\bf D}_{64}$ model, with the
anomaly-free spectrum  also found in \cite{gpdp}.


\begin{table}
\begin{center}
\begin{tabular}{|c|c|c|c|c|c|c|} 
\hline
Mod. & $n^{c}_{T}$ & $n^{c}_{H}$ & CP Gauge Group & 
$n^{o}_{V}$ & Charged Hypermultiplets & $n^{o}_{H}$ 
\\
\hline
\hline
${\bf A}_{64}$ & 5 & 16 & $Sp (8) \otimes Sp(8)$ & 72 & $({\bf 28},{\bf 1})
\oplus ({\bf 1},{\bf 28}) \oplus 2 ({\bf 8},{\bf 8})$ & 184 
\\
\hline
${\bf B}_{64}$ & 7 & 14 & $Sp (8) \otimes Sp(8)$ & 72 & $2 ({\bf 8},{\bf 8})$ &
128  
\\
\hline
${\bf C}_{64}$ & 9 & 12 & $U(8) \otimes U(8)$ & 128 & $({\bf 8},{\bf 8}^*) 
\oplus ({\bf 8}^*,{\bf 8})$ & 128 
\\
\hline
${\bf D}_{64}$ & 1 & 20 & $Sp (16) \otimes Sp(16)$ & 272 & $({\bf 120},{\bf 1})
\oplus ({\bf 1},{\bf 120}) \oplus ({\bf 16},{\bf 16})$ & 496 
\\
\hline
${\bf E}_{64}$ & 9 & 12 & $U(8)$ & 64 & ${\bf 64}$ & 64 
\\
\hline
${\bf F}_{64}$ & 9 & 12 & $U(4) \otimes U(4)$ & 32 & $({\bf 16},{\bf 1}) \oplus
({\bf 1},{\bf 16})$ & 32 
\\
\hline
${\bf G}_{64}$ & 9 & 12 & $U(4)$ & 16 & ${\bf 16}$ & 16 
\\
\hline
\end{tabular}
\end{center}
\caption{Open descendants of $(k=2)^4$ $Z_4$ Gepner model.}
\end{table}

\subsection{The $(k=1)^6$ Models}

The last class of models that we would like to discuss descends from a $Z_3$
orbifold of the maximal torus of $SU(3)\otimes SU(3)$,
or equivalently \cite{eoty} from the tensor product of
six copies of the $k=1$ minimal $N=2$ superconformal theory with $c=1$.
The latter is equivalent to a free boson at radius $R=\sqrt{12}$.
The twelve primary fields have conformal weights
$h_m=m^2/24$ with $m=-5,...,6$, and the corresponding characters will be
denoted by $\xi_m$. 

The graviton orbit \cite{eoty} that gives rise to the identity
character
\be
N^{+}_{0} = (\xi_0 + \xi_6 )^6 + (\xi_2 +\xi_{-4})^6 + (\xi_4 + \xi_{-2})^6
\ee
is the only self-conjugate one. Resolving the fixed point ambiguity, one finds
20 more massless characters, and two sets of 30 characters with $h=5/6$
and $h=13/6$.

For the diagonal model (${\bf D}_{81}$), the unoriented closed
spectrum resulting from a standard Klein bottle projection
includes the supergravity multiplet and 
one hypermultiplet from the identity character.
The other 20 massless characters give half hypermultiplet each. The resulting
spectrum contains {\it no tensor multiplets}, and
the dilaton must thus lie in a hypermultiplet. 
This closed spectrum is anomalous and requires the introduction of 
open strings. Tadpole cancellation in the transverse
channel selects the CP gauge group $SO(8)$ and,
aside from the vector multiplet, there are 10 hypermultiplets 
in the adjoint representation.
This representation is anomaly free and the chiral
fermion content of the open-string spectrum exactly cancels the gravitational
anomaly in the closed-string spectrum. 
Upon reduction on $T_2$ to $D=4$, the theory would not
be asymptotically free. It would be very interesting to find a model with no
tensor  multiplets with an asymptotically free CP group.
The analytic prepotential of the 
resulting effective field theory would not receive spacetime instanton
corrections.

The charge conjugation modular invariant (${\bf C}_{81}$)
leads to an open descendant with 10 hypermultiplets and
10 tensor multiplets arising from massless characters different from the identity, 
and one hypermultiplet and the supergravity multiplet arising from the identity.
The closed spectrum is anomalous and tadpole cancellation requires the 
introduction of open strings. The gauge group is once again 
$SO(8)$, but the spectrum contains only the vector multiplet. The open spectrum
differs from the one found in \cite{gpdp,gjdp} for the $Z_3$ orbifold, since the
Gepner model involves a rank 6 NS-NS antisymmetric tensor background.

An exhaustive analysis of all modular invariant torus amplitudes results in six
equivalent type IIB parent theories. The type I spectra are summarized in table
3 for some simple choices of the CP multiplicities. Although it is not evident 
from the table, $SO(8)$ is always allowed. This is precisely
the subgroup of $SO(32)$ preserving a generic configuration of 24 instantons on
K3 as required by anomaly cancellation, and suggests an F-theory interpretation
for these models. For an elliptic CY 3-fold $X$ fibered over a base $B$, the
number of tensor multiplets is $n_T=h_{11}(B)-1$, the rank of
the gauge group is 
$r_V=h_{11}(X)-h_{11}(B)-1$ and the number of neutral hypermultiplets is
$n^0_H=h_{12}(X)+1$ \cite{morvaf}.
A large class of elliptic CY 3-folds, studied by
Voisin and Borcea and classified by Nikulin in terms of three invariants
$(r,a,\delta )$, have Hodge numbers $h_{11} (X) = 5 + 3r -a$ and
$h_{12} (X) = 65 - 3r - 2a$. F-theory compactification on these spaces leads
to an $SO(8)^{k+1}$ gauge group, with
$k=(r-a)/2$, and to $g=(22-r-a)/2$ adjoint hypermultiples. It is
tempting to conjecture that the type I models with $G=SO(8)$
correspond to the choice $r=a=n_{T}^{c} + 1$.


\begin{table}
\begin{center}
\begin{tabular}{|c|c|c|c|c|c|c|} 
\hline
Mod. & $n^{c}_{T}$ & $n^{c}_{H}$ & CP Gauge Group & 
$n^{o}_{ V}$ & Charged Hypermultiplets & $n^{o}_{H}$ 
\\
\hline
\hline
${\bf A}_{81}$ & 6 & 15 & $U(4)$ & 16 & $4\times {\bf 16} \oplus 3\times ({\bf 6} 
\oplus {\bf 6}^* )$ & 100
\\
\hline
${\bf B}_{81}$ & 8 & 13 & $U(4)$ & 16 & $2\times {\bf 16} \oplus {\bf 6} 
\oplus {\bf 6}^* $ & 44
\\
\hline
${\bf C}_{81}$ & 10 & 11 & $SO(8)$ & 28 & --- & 0
\\
\hline
${\bf D}_{81}$ & 0 & 21 & $SO(8)$ & 28 & $10\times {\bf 28}$ & 280
\\
\hline
${\bf E}_{81}$ & 10 & 11 & $SO(8)$ & 28 & --- & 0
\\
\hline
${\bf F}_{81}$ & 10 & 11 & $SO(8)$ & 28 & --- & 0
\\
\hline
\end{tabular}
\end{center}
\caption{Open descendants of $(k=1)^6$ Gepner model.}
\end{table}

\section{Anomaly Cancellation and Final Remarks} 

Aside from $U(1)$ anomalies, tadpole cancellations guarantee the absence
of gauge and gravitational anomalies for all the models discussed so far.
Indeed, the antisymmetric tensor fields that flow in the transverse
channel take part in a generalized GS mechanism \cite{tensor} that allows the
cancellation of not necessarily factorized 6D chiral anomalies.
In table 4 we report the anomaly polynomials for the models discussed
in the previous section. These expressions determine the kinetic 
terms of the gauge fields and their singularities at finite
coupling \cite{tensor}, that have been associated to phase 
transitions \cite{dmw} with tensionless
strings \cite{blpssw}. Antisymmetric tensor fields, however,
can not cancel chiral anomalies
for abelian gauge groups with charged hypermultiplets. 
It has been proposed \cite{fms,blpssw} that a mechanism
similar to the one taking place in D=4 be at work in this case. In six
dimensions the relevant field
is a 4-form potential, dual to a pseudo-scalar. The massless
type I spectra that we have discussed include several R-R scalars.
These admit exact Peccei-Quinn symmetries
and may be dualized into 4-form potentials, that can couple to abelian fields 
via $A_4\wedge dA_1$. In the presence
of chiral anomalies, these R-R fields could then provide the
longitudinal degrees of freedom needed to exclude the (massive) 
abelian vector fields from the low-energy spectrum, much in the same way as in
$D=4$ \cite{dsw,wroma2}. 

\begin{table}
\begin{center}
\begin{tabular}{|c|l|} 
\hline
Model & Anomaly Polynomial 
\\
\hline
\hline
${\bf A}_{16}$ & $- {1\over 16} ({1\over 2} {\rm tr}R^2
 - {\rm tr} F_{1}^{2} - {\rm tr} F_{2}^{2} - {\rm tr} F_{3}^{2} -
{\rm tr} F_{4}^{2} )^2 $
\\
 & $+ {1\over 8}  ({\rm tr} F_{1}^{2} + {\rm tr} F_{2}^{2} - {\rm tr}
F_{3}^{2} - {\rm tr} F_{4}^{2} )^2 + 
{1\over 16} ({\rm tr} F_{1}^{2} - {\rm tr} F_{2}^{2} +
{\rm tr} F_{3}^{2} - {\rm tr} F_{4}^{2} )^2 $
\\
\hline
${\bf D}_{16}$ & $- {1\over 32} ({\rm tr} F_{1}^{2} + {\rm tr} F_{2}^{2} +
{\rm tr} F_{3}^{2} + {\rm tr} F_{4}^{2} + {\rm tr} R^2 )^2
+ {3 \over 32} ({\rm tr} F_{1}^{2} - {\rm tr} F_{2}^{2} -
{\rm tr} F_{3}^{2} + {\rm tr} F_{4}^{2} )^2$
\\
 & $+ {1\over 32} ({\rm tr} F_{1}^{2} + {\rm tr} F_{2}^{2} -
{\rm tr} F_{3}^{2} - {\rm tr} F_{4}^{2} )^2
+ {1\over 32} ({\rm tr} F_{1}^{2} - {\rm tr} F_{2}^{2} +
{\rm tr} F_{3}^{2} - {\rm tr} F_{4}^{2} )^2$
\\
\hline
${\bf A}_{64}$ & $-{1\over 8} ({1\over 2} {\rm tr} R^2 - {\rm tr} F_{1}^{2}
- {\rm tr} F_{2}^{2} )^2 + {1\over 8} ({\rm tr} F_{1}^{2}
- {\rm tr} F_{2}^{2} )^2$
\\
\hline
${\bf B}_{64}$ & $-{1\over 16} ({1\over 2} {\rm tr} R^2 - {\rm tr} F_{1}^{2}
- {\rm tr} F_{2}^{2} )^2 + {3\over 16} ({\rm tr} F_{1}^{2}
- {\rm tr} F_{2}^{2} )^2$
\\
\hline
${\bf C}_{64}$ & ${1\over 4} ({\rm tr} F_{1}^{2} - {\rm tr} F_{2}^{2} )^2
 + {1\over 3} ({\rm tr} F_{1} - {\rm tr} F_{2}) [ 
({\rm tr} F_{1}^{3} - {\rm tr} F_{2}^{3}) - {1\over 16} {\rm tr} R^2
({\rm tr} F_{1} - {\rm tr} F_{2}) ]$
\\
\hline
${\bf D}_{64}$ & $- {1\over 16} ({\rm tr} R^2 - {\rm tr} F_{1}^{2} - 
{\rm tr} F_{2}^{2} )^2 + {1\over 16} ({\rm tr} F_{1}^{2} - 
{\rm tr} F_{2}^{2} )^2$
\\
\hline
${\bf A}_{81}$ & $- {3\over 8} ( {1\over 4} {\rm tr} R^2 + 2 {\rm tr} F^2 )^2
+ {1\over 8} ({\rm tr} R^2 {\rm tr} F - 16 {\rm tr} F^3 ) {\rm tr} F$
\\
\hline
${\bf B}_{81}$ & $- {1\over 8} ( {1\over 4} {\rm tr} R^2 - 2 {\rm tr} F^2 )^2
+ {1\over 24} ({\rm tr} R^2 {\rm tr} F - 16 {\rm tr} F^3 ) {\rm tr} F$
\\
\hline
${\bf C}_{81}$ & $ {1\over 8} ({1\over 4} {\rm tr} R^2 - {\rm tr} F^{2})^2$
\\
\hline
${\bf D}_{81}$ & $-{9\over 8} ({1\over 4} {\rm tr} R^2 - {\rm tr} F^2 )^2$
\\
\hline
${\bf E}_{81}$ & ${1\over 8} ({1\over 4} {\rm tr} R^2 - {\rm tr} F^2 )^2$
\\
\hline
${\bf F}_{81}$ & ${1\over 8} ({1\over 4} {\rm tr} R^2 - {\rm tr} F^2 )^2$
\\
\hline
\end{tabular}
\end{center}
\caption{Non-vanishing anomaly polynomials for the models in tables 1,2,3, 
with $F_i$
the field strength of the $i$-th factor in the CP group. $R$ is the
curvature 2-form, and tr denotes the trace in the fundamental representation.}
\end{table}

The list of 6D Gepner models is clearly not exhausted by these
simple  cases, and a more complete analysis will be reported in \cite{abpss}.
We have also constructed the simplest 4D Gepner model with $N=1$
supersymmetry,
$(k=1)^9$. Since it is related to the $Z_3$ orbifold of the maximal torus of
$SU(3)^{\otimes 3}$, the CP multiplicities are reduced \cite{bpstor} by 
the non-vanishing (quantized) NS-NS antisymmetric tensor background.
The spectrum is encoded in 2187 characters. 
Apart from the identity, only 168 of them are massless. 
There are many torus amplitudes that give rise to different descendants.
In particular, the type IIB theory based on the charge 
conjugation invariant may be related to the compactification on a CY threefold 
with Hodge numbers $h_{11}=84$ and $h_{21}=0$. The open descendant has
$84+1$ chiral multiplets in the unoriented closed spectrum and an $Sp(4)$ CP group
with 84 chiral multiplets in the adjoint representation in the open 
spectrum. In the diagonal case one finds the same closed
spectrum and CP group without charged matter, while the
type IIB spectrum, with 84 $N=2$ vector multiplets and one universal 
hypermultiplet, may be related to the mirror of the above threefold.

\end{document}